\documentstyle[12pt]{article}

\catcode`\@=11

\def\section{\@startsection{section}{1}{\z@}{3.5ex plus 1ex minus
 .2ex}{2.3ex plus .2ex}{\bf}}

\def\thesubsection{\arabic{section}.\arabic{subsection}}
\renewcommand{\subsection}[1]{\addtocounter{subsection}{1}
\vspace{2.5mm}\par\noindent {\it \thesubsection . #1}\par
 \vspace{0.5mm} }
\catcode`\@=12
\newfont{\mbm}{msbm10 scaled\magstep1}

\def\reflist{\section*{References\markboth
        {REFLIST}{REFLIST}}\list
        {[\arabic{enumi}]\hfill}{\settowidth\labelwidth{[999]}
        \leftmargin\labelwidth
        \advance\leftmargin\labelsep\usecounter{enumi}}}
 \relax

\newcommand{\be}{\begin{equation}}
\newcommand{\ee}{\end{equation}}
\newcommand{\ba}{\begin{eqnarray}}
\newcommand{\ea}{\end{eqnarray}}

\catcode`\@=11

\begin{document}
\begin{titlepage}
\rightline{{CPHT RR 068.0902}}
\rightline{{LPT-ORSAY 02-136}}

\vskip 2cm
\begin{center}{{\huge \bf 
Time and space dependent backgrounds
from nonsupersymmetric strings }}
\end{center}
\vskip 1cm
\centerline{E. Dudas${}^{\dagger,\star}$,
J. Mourad${}^{\dagger,*}$ and C. Timirgaziu${}^{\dagger}$}
\vskip 0.5cm
\centerline{\it ${}^\dagger$ Laboratoire de Physique 
Th{\'e}orique
\footnote{Unit{\'e} Mixte de Recherche du CNRS (UMR 8627).}}
\centerline{\it Universit{\'e} de Paris-Sud, B{\^a}t. 210, F-91405 Orsay Cedex}
\vskip 0.3cm 
\centerline{\it ${}^\star$ Centre de Physique Th{\'e}orique,
Ecole Polytechnique, F-91128 Palaiseau}
\vskip 0.3cm 
\centerline{\it ${}^*$ F{\'e}d{\'e}ration de Recherche APC,
Universit{\'e} de Paris 7}

\vskip  1.0cm
\begin{abstract}

We investigate  maximally symmetric backgrounds in 
nonsupersymmetric string
vacua with D-branes and O-planes localized in the compact space. 
We find a class of solutions with a perturbative
string coupling constant in all regions of spacetime.
Depending on the particular model, we find either a 
time evolution with  
a big-bang type singularity or a space dependent background
with generically orbifold singularities.
 We show that the result can be interpreted as a  
supersymmetric bulk with  
some symmetries broken by the boundaries. We also discuss an
interesting connection to Lorentzian and Euclidian orbifolds. 
\end{abstract}
\end{titlepage}

\section{Introduction and summary of results}

String theory provides a natural
setting to address cosmological issues like
the cosmological constant problem or the fate of
the big-bang singularity. 
Whereas the first problem still awaits for a qualitatively
different perspective, the second one led over the last ten years to
more explicit proposals like 
the pre-big bang model \cite{veneziano}, the  ekpyrotic scenario 
\cite{epkyrotic} 
or brane-world models \cite{world}. In the second model, objects with 
negative tensions were important for its realisation.
In addition to the positive tension branes, orientifolds
allow also for negative tension objects: the O-planes. This
renders them candidates for rich possibilities of
cosmological backgrounds. The goal of this
paper is to study these possibilities.

The  string models we consider   
are vacua with D-branes and orientifold planes,
\cite{augusto1,Polchinski:1996na} and with broken supersymmetry. 
Orientifold models with
D-branes/O-planes with broken supersymmetry in
various sectors of the theory were proposed in the last
few years \cite{bd,ads1,sugimoto,bsb}. The classical background
of such vacua has typically lower symmetry 
and was explicitly worked out in some particular examples \cite{dm1}.
An obvious and important worry about such constructions is the issue
of classical and quantum stability and their fate. The fact that some of
these constructions \cite{sugimoto,bsb} are tachyon-free in all moduli 
space of the
theory is a promising feature, but gives only a partial answer to the
stability question. On the other hand, it is clear that typically, as
soon as supersymmetry is broken, D-branes/O-planes start to interact,
curve the internal or the noncompact space and 
generically produce a
time-dependent configuration. 
 
The explicit models we
consider are nine dimensional string models. At the effective field
theory level, supersymmetry breaking is described by 
non-BPS configurations of D-branes and O-planes, as
as well as a one-loop bulk cosmological constant. 
 
The maximally symmetric classical background of these
models generically depends on two coordinates, 
which, according to the
details of the models, can be $(t,y)$ or $(z,y)$, where $t$ is the
time, $y$ is the compact coordinate orthogonal to the branes and $z$
is a noncompact coordinate parallel to the branes. 
We concentrate on solutions which
are perturbative in the string coupling constant so
that the classical solution receives small quantum corrections.
In the time dependent case, we show by an adequate choice
of coordinates that one of the solutions 
has a static bulk and hence has as much symmetries as the
supersymmetric bulk, but the boundaries are moving
and thus break the invariance under time translation.
The solutions are characterized by a big bang (or big crunch)
singularity which is due to
the collision of the two boundaries. Interestingly,
this solution can be interpreted also as an
orbifold by a boost of a static and supersymmetric 
background.
The boost parameter on the Lorentzian side is related to the branes
and O-planes and the one-loop cosmological constant. We also
show that in models with NS-NS tadpoles, the usual NS-NS
tadpole condition is replaced by a sum rule,
which is basically the boosted version of the static tadpole condition.
In the space dependent case, the boundaries join at a
conical singularity and also break some of the bulk symmetries.
We show that this background can be considered as an orbifold
by a two dimensional rotation of the supersymmetric one.  

One application of our work is to the big bang type cosmology. In
this respect, our results have similar features to the pre-big bang 
\cite{veneziano} and ekpyrotic \cite{epkyrotic} scenarios. 
Another possible application
is to the Fischler-Susskind mechanism. The relation we find between the
classical solutions of nonsupersymmetric orientifolds and lorentzian
orbifolds suggests a deepest relation at the quantum level.
In this respect, a severe instability of lorentzian orbifolds was 
recently pointed in \cite{hp} along with some possible ways out 
\cite{lms,hp}. Irrespective
of the final fate of lorentzian orbifolds, we believe that connections between
seemingly unrelated vacua can be useful for a better
understanding of perturbative and nonperturbative aspects of string theory.
We hope to come back to this issue in the future.

The paper is organized as follows. 
Section 2 describes the various classical and perturbative
solutions for string models with D-branes and O-planes in nine
dimensions and their relation to Lorentzian and Euclidian
orbifolds. Section 3 discusses some of their 
applications to the big bang and big crunch cosmology, whereas 
Section 4 presents some explicit string examples. 
Appendix A contains the technical 
details involved in the resolution of the equations
of motion and a more complete set of various classical
solutions, including the ones with strong string coupling in some
regions of spacetime. Appendix B gives some details on the explicit string
models under consideration. 
             
\section{Time and space-dependent backgrounds of nonsupersymmetric strings}

We consider a generic form of the effective action of the Type II string
containing D8 branes (and eventually O8 planes in the case of
orientifold models). We include also the bulk one-loop cosmological constant
$\Lambda_1$ which typically arise in
most nonsupersymmetric strings. The resulting effective action in the
string frame reads

\ba
S &=& {1 \over 2{\kappa}^2} \int d^{10} x \sqrt {-G} \biggl[ e^{-2 \Phi} ( R- {1
\over 2} ({\partial \Phi})^2)- {1 \over 2 \times 10 !} F_{10}^2 -2 {\kappa}^2 \Lambda_1 
\biggr] \nonumber \\
&-& \int_{y=0} d^9 x (T_0 \ \sqrt{-\gamma} \ e^{-\Phi}
\ + q_0 \ A_9) - \int_{y= \pi R} d^9 x  (T_1 \sqrt{-\gamma} \ e^{-\Phi}
\ +q_1 A_9 ) \ , \label{t1b}
\ea
where $\Phi$ is the dilaton, $A_9$ the RR nine-form coupling to D8
branes and O8 planes and $\gamma$ is the induced metric. For
simplicity of the discussion we placed all branes either at the origin
$y=0$ or at $y=\pi R$ of a compact coordinate $y$ of radius $R$.
Since supersymmetry is broken, we do not assume any relation between the RR charges $q_i$ 
and the NS tensions $T_i$.

We consider in the following string vacua of the type \cite{sugimoto},
\cite{bsb}, where the RR charge is globally cancelled, as required by
consistency arguments, but the NS-NS tadpole conditions are violated
and there is an induced one-loop cosmological constant
\ba
q_0 \ + q_1 &=& 0 \quad ({\rm RR \ tadpole \ conditions}) \ , \nonumber \\
T_0 \ + T_1 &\not= 0& \quad ({\rm uncancelled \ NS-NS \ tadpoles }) \ 
,  \nonumber \\
\Lambda_1 &\not=& 0 \ . \ \label{t1}
\ea
The classical field equations have no solution with $SO(9)$ 
symmetry, in agreement with various arguments 
presented in the literature \cite{dm1}.  
We search here for solutions depending on the compact
coordinate $y$ and on another coordinate, which can be the time $t$ or
another space coordinate $z$. We restrict ourselves in that section 
to solutions which
are smoothly connected to the supersymmetric ones \cite{pw} and 
have a perturbative string coupling throughout the
spacetime. We leave to the Appendix A the explicit derivation of
the  complete set of classical solutions.

\subsection{Cosmological solutions}

The general form of such a solution is of the form
\ba
ds^2 &=& e^{2 A (t,y)} \left(1+{k x^2 \over
4}\right)^{-2} \delta_{\mu \nu} dx^{\mu} dx^{\nu} + 
e^{2 B (t,y)} (-dt^2 + dy^2) \ , \nonumber \\
F_{10} &=& f(t,y) \ \epsilon_{10} \quad , \quad \Phi = \Phi (t,y) 
\ , \label{t2}  
\ea 
where $\epsilon_{10}$ is the ten-dimensional volume form.
The eight-dimensional metric at fixed $y$ and $t$ is a maximally
symmetric one: $k=0$ for a flat eight dimensional space, $k=1$ for a closed
8-sphere and $k=-1$ for an open 8-hyperboloid.
 
The equations of motion of the nine-form can be readily solved 
and the solution, in the Einstein frame, reads
\be
f=-q_0 \ {\kappa}^2 e^{5\Phi/2} \ \epsilon (y) \ , \label{t03}
\ee
where  
$\epsilon(y)$ is an odd $2\pi R$-periodic function 
and $\epsilon(y)=1$ when $y$ is between $0$ and $\pi R$. Since most of
our results ask for the existence of negative tension objects, we mostly
refer in the following to orientifold models \cite{augusto1} and
consequently we choose the ten-form to be odd under $y \rightarrow - y$. 
Since the nine-form potential has no physical degrees of freedom, we
replace it in the Lagrangian by its classical expression (\ref{t03}). 
As a result, we get an effective one loop cosmological
constant given by
\be
\Lambda_e \ = \ \Lambda_1+{q_0^2 \over 4} {\kappa}^2 \ . \label{t04}
\ee
We define for later convenience 
$\beta_0 = -T_0 {\kappa}^2$ and $\beta_1= T_1 {\kappa}^2$.
The Einstein and dilaton equations for the remaining functions are 
given by\footnote{The
notations we use are $\Phi_{tt} = {\partial^2 \Phi / \partial t^2}
$, etc.}
\ba
&&-\Phi_{tt}+\Phi_{yy}-8A_t\Phi_t+8A_y\Phi_y-{5 \kappa^2 \Lambda_e
}
e^{2B+5\Phi/2}=-{5 \over
2}S \ , \label{e1} \\
&& -7A_{tt}+7A_{yy}-28A_t^2+28A_y^2-B_{tt}+B_{yy}
-{\Phi_t^2\over 4}+{\Phi_y^2\over 4}-21 k e^{2B-2A}+
{\kappa^2 \Lambda_e }e^{2B+5\Phi/2}
=S,\label{e2} \\
&& -8A_{tt}-36A_t^2+28A_y^2+8A_tB_t+8A_yB_y-
{\Phi_t^2\over 4}-{\Phi_y^2\over 4}-
28 k e^{2B-2A}+{\kappa^2 \Lambda_e }
e^{2B+5\Phi/2}=0,\label{e3} \\
&&-8A_{yy}+28A_t^2-36A_y^2+8A_tB_t+8A_yB_y-
{\Phi_t^2\over 4}-{\Phi_y^2\over 4}+28
 k e^{2B-2A}-{\kappa^2 \Lambda_e}
e^{2B+5\Phi/2}=-S \label{e4} \ , \\
&&A_{ty}+A_tA_y-A_tB_y-A_yB_t+{\Phi_t\Phi_y \over 16}=0\label{e5} \ ,
\ea
 where we have defined 
\be
S = e^{5\Phi/4+B} \ [\beta_0\delta(y)-\beta_1\delta(y-\pi R)] \ .
\ee

All the functions $A$, $B$ and $\Phi$ are even in $y$ and $2\pi R$-
periodic. The sources in the right-hand side of the equations determine
the $y$ derivatives of these functions at $0$ and $\pi R$
as
\be
A_y(0^+,t)={\beta_0 \over 16} \ e^{B(0,t)+5\Phi(0,t)/4} \quad , \quad
A_y(\pi R^-,t)={\beta_1 \over 16} \ e^{B(\pi R,t)+5\Phi(\pi R,t)/4} \ ,
\label{bou}
\ee
and
\ba 
&&B_y(0^+,t)=A_y(0^+,t),\ \ \Phi_y(0^+,t)=-20 A_y(0^+,t) \ , \nonumber \\
&&B_y(\pi R^-,t)=A_y(\pi R ^-,t),\ \ \Phi_y(\pi R^-,t) =
-20 A_y(\pi R^-,t) \ . \label{boun}
\ea

The boundary conditions (\ref{boun})
imply the relations:
\be
\Phi(y,t) = - 20 A(y,t)+\phi(t) \quad , \quad B(y,t)=A(y,t)+b(t) \ , \label{an}
\ee
where $\phi$ and $b$ are for the time being arbitrary functions of time.
The relations (\ref{an}) have the virtue of reducing the boundary conditions
to (\ref{bou}). Using the relations (\ref{an}) in equation (\ref{e5}),
one gets
\be
[e^{24 A-b-5\phi/4}]_{yt} \ = \ 0 \ ,
\ee
which is readily solved by
\be
e^{24 A-b-5\phi/4} \ = \ F(y)+G(t) \ . \label{fac}
\ee
The relations (\ref{an}) and (\ref{fac}) allow to 
transform the partial differential equations (\ref{e1}-\ref{e4})
into ordinary differential equations for $b$, $\phi$, $F$ and $G$.
The boundary conditions (\ref{bou}) translate into
\be
F'(0^+)={3\beta_0 \over 2} \quad , \quad F'(\pi R^-) = {3\beta_1 \over
2} \ . \label{bor}
\ee

The  equations resulting from the
substitution in  (\ref{e1}-\ref{e4})
of the relations (\ref{an}) and (\ref{fac})
are discussed in some details in Appendix A.
We show there that for $k=0$ it is possible to
solve exactly and to find all the solutions of the equations of
motion. There are two classes of
solutions. The first one, denoted {\bf a} in the Appendix A, is 
characterized by
\ba
&&{\dot \phi}=0 \quad , \quad {\dot b}^2=\lambda^2 \quad , \quad
{\dot G} {\dot b}+ \lambda^2 G=0 \ , \label{eq1}\\
&&F'^2-\lambda^2F^2={9 \kappa^2 \Lambda_e  } \ , \label{eq}
\ea
where $\lambda$ is a positive constant. In (\ref{bor}), (\ref{eq1}) and
(\ref{eq}) the prime (the dot) denote differentiation with respect to 
$y$ ($t$). There is a second class of solutions, called {\bf b} in the
Appendix, which are characterized by $G=0$. It is shown in the Appendix
that the solutions in the first class  are the only ones with a
perturbative value of the string coupling in the whole spacetime.
In addition, they are the only ones which are smoothly connected to the
supersymmetric solutions \cite{pw}. Notice the remarquable fact that 
eqs. (\ref{eq1}) and (\ref{eq}) are of first order, even though
they are not derived from BPS-type conditions. 

The boundary conditions are not always compatible with
equations (\ref{eq}). Evaluating (\ref{eq}) at the origin
and at $\pi R$ we get the two conditions
\be
T_0^2 \geq q_0^2 + 4 {\Lambda_1 \over {\kappa}^2} \quad , \quad 
T_1^2 \geq q_1^2 + 4 {\Lambda_1 \over {\kappa}^2} \ . \label{condi}
\ee
These conditions are necessary but, as we will see, not
sufficient to insure the existence of a solution in this class.
Before analysing in detail 
the solutions for the different models, notice that if the
effective cosmological constant $\Lambda_e$ is negative
eqs. (\ref{condi}) are automatically satisfied. 
This is to be compared to the sum rules 
\cite{gibbons} where staticity and compactness 
impose severe restrictions (or fine tuning) on the tensions.

The SUSY configuration is a special case of the 
above system: it corresponds to a vanishing one loop
cosmological constant, $\Lambda_1=0$,
and to the equality of the RR charges and NS charges
$q_0 = T_0 = - T_1$. This implies that $\lambda=0$, $F'$ is constant 
and the space is flat, $k=0$.
 
If neither of the two conditions (\ref{condi}) are satisfied then
the only solutions for the background of the form (\ref{t2}) have
singularities in the string coupling and are displayed for completeness in the Appendix. 
However, a non-singular solution with the same number
of isometries exists, it amounts to interchange the time 
coordinate with one of the
eight coordinates $x$, called $z$ later on. If one the conditions(\ref{condi})
is true but not the other, then one has to look for a solution  with lower symmetries.
There are three qualitatively distinct cases to consider, depending on
the value of the effective one-loop cosmological constant (\ref{t04})~: 

\begin{itemize}

\item{\bf i)} If $\Lambda_e > 0 $, 
then the solution has the form\footnote{There is also the solution
$F \rightarrow -F$ in (\ref{t6}). This is equivalent, however, to a
reflection $y \rightarrow -y + \pi R$ which exchanges the
two fixed points, accompanied by the replacement $\omega \rightarrow
-\omega - \pi \lambda R$.}
\be
F(y) = {3 \kappa \sqrt{\Lambda_e} \over \lambda}
sh(\lambda |y|+\omega) \ , \label{t6}
\ee
where the boundary conditions (\ref{bor}) determine the parameters
$\lambda$ and $\omega$ 
\be
ch(\omega)= -T_0 {\kappa} /(2\sqrt{\Lambda_e}) \quad , \quad
ch(\pi \lambda R+\omega)= T_1 {\kappa} / (2\sqrt{\Lambda_e}) \ . \label{t7} 
\ee
Notice that (\ref{t7}) can have a solution only if the tensions in $y=0$
and $y=\pi R$ have opposite signs. Therefore we need objects of
negative tension in the theory, which in our explicit string examples later on
are orientifold planes. 

The final solutions of the classical field equations, 
in the Einstein frame, read
\ba
e^{24 A} &=& e^{b_0+5 \phi_0/4} \biggl[G_0 + {3 {\kappa} \sqrt{\Lambda_e} \over \lambda} e^{
\lambda t} sh (\lambda |y|+\omega) \biggr] \ , \nonumber \\
e^{24 B} &=&   e^{24 \lambda t +25 b_0+5 \phi_0/4}
\biggl[G_0+ {3 {\kappa} \sqrt{\Lambda_e} \over \lambda} e^{
\lambda t } sh (\lambda |y|+\omega) \biggr] \ , \nonumber \\
e^{\Phi} &=& e^{-5 b_0/6 - \phi_0/24} \biggl[G_0 + {3 {\kappa} \sqrt{\Lambda_e} \over \lambda} e^{
\lambda t } sh (\lambda |y|+\omega) \biggr]^{-{5 \over 6}}  
 \ . \label{t3}
\ea
We did carefully keep track in (\ref{t3}) of the integration constants 
$\phi_0$, $b_0$ and $G_0$ . Notice that for $G_0 < 0$, there are
singularities in the $(t,y)$ plane. We restrict in this section to safer values $G_0 \geq 0$.
Notice that for $T_1 < |T_0|$ ,
we get $\omega <0$ and therefore there are
singularities if 
$G_0+ ({3 {\kappa} \sqrt{\Lambda_e} / \lambda}) \exp (
\lambda t ) \ sh (\omega) \leq 0$, whereas  for $T_1 > |T_0|$ ,
$\omega$ is positive and therefore there are no singularities in the
compact space. In the limit of vanishing $\lambda$ and
$\omega$, we recover the supersymmetric solution \cite{pw}.

Let us consider in more details 
the resulting space-time metric and
choose for simplicity $b_0=0$, $\phi_0=0$. We find
\be
ds^2 =  \biggl[ G_0+ {3 {\kappa} \sqrt{\Lambda_e} \over \lambda} e^{
\lambda t} sh (\lambda |y|+\omega) \biggr]^{1 \over 12} \ \biggl[  
\delta_{\mu \nu} dx^{\mu} dx^{\nu} + e^{2 \lambda t} (-dt^2+dy^2) \biggr]
\ . \label{t5}   
\ee
Finally, by making the change of variables
\be
T \ = {1 \over \lambda} \ e^{\lambda t} \ ch (\lambda y+\omega) \quad , \quad 
X \ = {1 \over \lambda} \ e^{\lambda t} \ sh (\lambda y+\omega)  \ , \label{t8} 
\ee
we get the spacetime metric
\be
ds^2 =  
\biggl[ G_0+ 3{\kappa} \sqrt{\Lambda_e} X   \biggr]^{1 \over 12} \biggl[  
\delta_{\mu \nu} dx^{\mu} dx^{\nu} -dT^2+ dX^2 \biggr]
\ , \label{t9}   
\ee
when $y>0$. The $Z_2$ identification
$y\rightarrow -y$ is mapped in terms of the coordinates $(X,T)$
to a parity $\Pi_X$ times a boost ${\cal K}$ with a parameter $2\omega$. This means that
the orientifold operation acts in the $(T,X)$ plane as 
\be
\Omega' \ = \ \Omega \ \Pi_X \ {\cal K}_{2 \omega} \ . \label{t09}
\ee
In addition, the identification of points on the circle $y = y+2 \pi
R$ results in $(T,X)$ coordinates in the orbifold identification 

\ba 
\left( 
\begin{array}{c} 
T \\ 
X 
\end{array} 
\right)  
& \ \rightarrow \ & 
\left(
\begin{array}{cc} 
ch (2 \pi \lambda R) & sh (2 \pi \lambda R)  \\ 
sh (2 \pi \lambda R)  & ch (2 \pi \lambda R) 
\end{array}
\right)
\left(
\begin{array}{c}
T \\
X
\end{array}
\right) \ , \label{t10}
\ea
which is nothing but a two-dimensional boost ${\cal K}_{2 \pi \lambda R}$ with a velocity 
$v= th(2 \pi \lambda R)$ in the $(T,X)$ space.
 
The final result (\ref{t9}) is quite surprising. 
Indeed, (\ref{t9}) coincides with the spacetime metric 
obtained in \cite{pw} in the supersymmetric
Type I' string with N D8 branes at the origin $X=0$ of a compact coordinate of
radius $R$ and $32-N$ D8 branes at $X=\pi R$~!
The supersymmetric Polchinski-Witten solution  and our
non-supersymmetric solution appear to be two different orbifolds
of the same ten-dimensional background. More precisely, the SUSY solution uses the
translation group, while the non-SUSY one uses the two-dimensional Lorentz group.

The metric (\ref{t9}) and the identifications
(\ref{t09}), (\ref{t10}) allow a  simple physical 
interpretation of our configuration in the $(X,T)$ coordinates. 
Indeed, the fixed points of the two orientifold operations are
\ba
\Omega'~: \  X &=& th \omega \ T \nonumber \\
\Omega'~ {\cal K}_{2 \pi \lambda R} : \  X &=& th (\pi \lambda R+ \omega ) \ T \ . \label{t010}
\ea
Consequently, the negative tension O-planes and
the branes located at the origin move with a constant velocity
\be
v_0 = th \ \omega \ , \label{t12}
\ee
in the static background (\ref{t9}), whereas the positive
tension branes and O-planes at $y=\pi R$ move at a
constant velocity 
\be
v_1=th \ (\pi\lambda R+\omega) \ . \label{t13}
\ee
Moreover, the boundary conditions (\ref{t7}) encode the dynamics
of the two boundaries in the condition
\be
T_0 \sqrt{1-v_0^2} + T_1 \sqrt{1-v_1^2} = 0 \ . \label{t14} 
\ee
The interpretation of (\ref{t14}) is quite simple. 
In the supersymmetric case, the branes and O-planes are at rest and (\ref{t14})
reduces to the standard NS-NS tadpole condition $T_0+T_1=0$. 
In the case with broken supersymmetry, the NS-NS tadpoles are ``boosted'' according
to the velocity of the branes and O-planes in the background
(\ref{t9}). The boost is the one appropriate for a lagrangian
density, since the dilaton field in (\ref{t1b}) couples to the
lagrangian, instead of the energy.

An interesting particular example of the above results is the one in
which all global tadpoles are
cancelled and we have a negative bulk cosmological constant
\be
T_0 +T_1 = 0 \ , \quad \Lambda_1 < 0 \ . \label{t014}
\ee
In this case it turns out that (\ref{t7}) is satisfied for
\ba
\omega &=& - \ {\pi \lambda R \over 2} \quad , \quad ch \ ({\pi \lambda R
\over 2}) \ = \ {T_1 {\kappa} \over 2 \sqrt{\Lambda_e}} \ , \nonumber \\
v_0 &=& - \ th \ ({\pi \lambda R \over 2}) \quad , \quad 
v_1 \ = \ th \ ({\pi \lambda R \over 2}) \ .  \label{t015}
\ea
Therefore even in the absence of disk NS-NS tadpoles, one-loop
cosmological constant is sufficient to generate a constant
velocity dynamics, which can be interpreted in
terms of orientifolds of a Lorentzian orbifold. Notice, however, that
in this case the metric and the dilaton can be singular between the O-planes. 
>From this perspective, models with
NS-NS tadpoles seem to be the only ones free of
singularities in the compact space.

\item{\bf ii)} If $\Lambda_e <0$,
then the conditions (\ref{condi}) are verified and
the solution for $F$ is given by 
\be
F(y)= \pm {3 \kappa \sqrt{- \Lambda_e} \over \lambda} \ 
 ch(\lambda |y|+ \omega) \ , \label{t16}
\ee
where $\omega$ and $\lambda$ are determined by 
\be
sh (\omega) = \mp T_0 {\kappa} /(2\sqrt{-\Lambda_e}) \quad , \quad
sh (\pi \lambda R+ \omega)= \pm T_1 {\kappa}/(2\sqrt{- \Lambda_e}) \ . \label{t17} 
\ee
The solution of the classical field equations reads
\ba
e^{24 A} &=& e^{b_0+5 \phi_0 /4} \biggl[ G_0 \pm {3 {\kappa}
\sqrt{-\Lambda_e} \over \lambda} e^{
\lambda t} ch (\lambda |y|+\omega) \biggr] \ , \nonumber \\
e^{24 B} &=&   e^{24 \lambda t+ 25 b_0+5 \phi_0 /4}
\biggl[ G_0 \pm {3 {\kappa} \sqrt{-\Lambda_e} \over \lambda} e^{
\lambda t} ch (\lambda |y|+\omega) \biggr] \ , \nonumber \\
e^{\Phi} &=&  e^{-5b_0/6- \phi_0 /24 } \biggl[ G_0 \pm {3 {\kappa}
\sqrt{-\Lambda_e} \over \lambda} e^{\lambda t } ch (\lambda |y|+\omega) 
\biggr]^{-{5 \over 6}} \ . \label{t18}
\ea 
By the same change of variables (\ref{t8}) and by setting for
simplicity $b_0 = \phi_0=0$, we get for $y > 0$
the spacetime metric
\be
ds^2 =  
\biggl[ G_0 \pm 3{\kappa} \sqrt{\Lambda_e} T  \biggr]^{1 \over 12} \biggl[  
\delta_{\mu \nu} dx^{\mu} dx^{\nu} -dT^2+ dX^2 \biggr]
\ . \label{t19}   
\ee
As in the case {\bf i)}, the $Z_2$ identification
$y\rightarrow -y$ is mapped in terms of the coordinates $(X,T)$
to a parity times a boost with a parameter $2\omega$,
while the identification of points on the circle $y = y+2 \pi
R$ gives for the $(T,X)$ coordinates the identification (\ref{t10}).
The boundary conditions (\ref{t17}) in this case imply the following
condition on the spacetime boundary velocities
\be
{T_0 \over v_0} \sqrt{1-v_0^2} + {T_1 \over v_1} \sqrt{1-v_1^2} = 0 \ . \label{t21} 
\ee
Notice that in this case it is possible to obtain solutions to
(\ref{t21}) with tensions of the same sign in the two fixed points,
provided that the two velocities (\ref{t12}), (\ref{t13}) have opposite signs.

\item{\bf iii)} If ${\Lambda}_e =0$,
then there are time-dependent solutions provided that the
tensions at $y=0$ and $y=\pi R$ have opposite signs. 
The final form of the solution is
\ba
ds^2 &=&  \biggl[ G_0 + F_0 \  e^{
\lambda (t\pm |y|)} \biggr]^{1 \over 12} \ \biggl[  
\delta_{\mu \nu} dx^{\mu} dx^{\nu} + e^{2 (\lambda t +b_0)} (-dt^2+dy^2) \biggr]
\ , \nonumber \\
e^{\Phi} &=& \ e^{\phi_0} 
 \biggl[ G_0 + F_0 \ e^{
\lambda (t\pm |y|)} \biggr]^{-{5 \over 6}}  \ , \label{t26}   
\ea

\end{itemize}

where $F_0$ is a constant and the parameter $\lambda$ in (\ref{t26}) is determined by
the condition
\be
e^{\pi \lambda R} = - {T_1 \over T_0} \ . \label{t27}
\ee
In this case, by introducing the coordinates (\ref{t8}) with $\omega=0$,
we find the spacetime metric
\be
ds^2 =  
\biggl[ G_0 + \lambda F_0 X^{\pm}  \biggr]^{1 \over 12} \biggl[  
\delta_{\mu \nu} dx^{\mu} dx^{\nu} -dT^2+ dX^2 \biggr]
\ , \label{t28}   
\ee
where we introduced the light-cone coordinates $X^{\pm}= T \pm X$. 
Notice from (\ref{t27}) that by taking $T_0$ very small and negative
$T_0 \rightarrow 0^{-}$, we can generate an infinite boost parameter $\lambda$.

\subsection{Static solutions}

One may wonder whether there are similar solutions 
which do not depend on time but on an additional 
space coordinate $z$, that is with the ansatz
\be
ds^2 = e^{2A} g_{\mu\nu} dx^\mu dx^\nu \ + \ 
e^{2B} (dz^2 + dy^2) \ , \label{t022}
\ee
where $g_{\mu \nu}$ is the eight dimensional flat, dS or AdS metric and
with $A$, $B$ and $\Phi$ functions of $y$ and $z$.
Here also we have two classes of solutions. The first one is also
characterized by a finite string coupling and is smoothly connected to
the supersymmetric solution \cite{pw}.
In this case, a similar analysis as before shows that the
eight-dimensional metric $g_{\mu \nu}$ is flat and, with the replacement
$t \rightarrow z$ equations (\ref{fac}), (\ref{bor}) and
(\ref{eq1}) hold true. The
nontrivial modification is a crucial sign in equation (\ref{eq})
which becomes
 \be
F'^2 + \lambda^2 F^2  = 9 {\kappa^2 \Lambda}_e \ . \label{eq43}
\ee
If we allow for a negative effective one loop cosmological constant,
we see at once that (\ref{eq43}) cannot be satisfied
and the time dependent solutions are the only ones.
Evaluating (\ref{eq43}) at the origin
and at $\pi R$ we get the two conditions
\be
T_0^2 \leq q_0^2 + 4 {\Lambda_1 \over {\kappa}^2} \quad , \quad 
T_1^2 \leq q_1^2 + 4 {\Lambda_1 \over {\kappa}^2} \ . \label{condi2}
\ee
Unlike in (\ref{condi}), these conditions are necessary and
sufficient to insure the existence of a solution.
The explicit form of the solution in this case is
\be
F(y) = {3 \kappa \sqrt{ \Lambda_e} \over \lambda}
\sin (\lambda |y|+\omega) \ , \label{t22}
\ee
where the boundary conditions (\ref{bor}) determine the parameters
$\lambda$ and $\omega$ 
\be
\cos (\omega)= -T_0 {\kappa} /(2\sqrt{\Lambda_e}) \quad , \quad
\cos (\pi \lambda R + \omega)= T_1 {\kappa} / (2\sqrt{\Lambda_e}) \ . \label{t23} 
\ee

The final solutions of the classical field equations, 
in the Einstein frame, read
\ba
e^{24 A} &=& e^{b_0+5 \phi_0/4} \biggl[G_0 + {3 {\kappa} \sqrt{\Lambda_e} \over \lambda} e^{
\lambda z} \sin (\lambda |y|+\omega) \biggr] \ , \nonumber \\
e^{24 B} &=&   e^{24 \lambda z +25 b_0+5 \phi_0/4}
\biggl[G_0+ {3 {\kappa} \sqrt{\Lambda_e} \over \lambda} e^{
\lambda z } \sin (\lambda |y|+\omega) \biggr] \ , \nonumber \\
e^{\Phi} &=& e^{-5 b_0/6 - \phi_0/24} \biggl[G_0 + {3 {\kappa} \sqrt{\Lambda_e} \over \lambda} e^{
\lambda z } \sin (\lambda |y|+\omega) \biggr]^{-{5 \over 6}}  
 \ . \label{t24}
\ea
This solution is continously connected to the supersymmetric solution
\cite{pw} in the limit of vanishing $\lambda$ and $\omega$. 

The analog of the equation (\ref{t14}) in this case is
\be
{T_0 \over \cos \omega} + {T_1 \over \cos (\lambda \pi R+ \omega)} = 0 \
. \label{t25} 
\ee
Notice that, contrary to the previous time-dependent case (\ref{t14}),
eq. (\ref{t25}) does not give any restriction on the tensions. 

The $z$ coordinate is noncompact and the Planck mass in this
background is infinite.
There are singularities for $z = \infty$ (or $z = - \infty$ for the
negative branch solution $z \rightarrow -z$). 
Depending on the sign of $G_0$ and the numerical values of $\lambda$
and $\omega$, this solutions can also have singularities at a finite
distance from the origin in the $(z,y)$ plane .

Interestingly, this solution can also be related to the supersymmetric
solution \cite{pw}. Indeed, by the change of coordinates 
\be
Y \ = {1 \over \lambda} \ e^{\lambda z} \ \sin (\lambda y + \omega) \quad , \quad 
Z \ = {1 \over \lambda} \ e^{\lambda z} \ \cos (\lambda y + \omega)  \ , \label{t025} 
\ee
we get for $y >0$ the spacetime metric
\be
ds^2 =  
\biggl[ G_0+ 3{\kappa} \sqrt{\Lambda_e} Y   \biggr]^{1 \over 12} \biggl[  
\eta_{\mu \nu} dx^{\mu} dx^{\nu} +dY^2+ dZ^2 \biggr]
\ , \label{t026}   
\ee
which is the one derived by Polchinski and Witten \cite{pw}, except that
here the $Y$ coordinate is noncompact.   
The periodicity $y = y+2 \pi R$ reflects in the new coordinate system
$(Z,Y)$ in the orbifold identification

\ba 
\left( 
\begin{array}{c} 
Z \\ 
Y 
\end{array} 
\right)  
& \ \rightarrow \ & 
\left(
\begin{array}{cc} 
\cos (2 \pi \lambda R) & - \sin (2 \pi \lambda R)  \\ 
\sin (2 \pi \lambda R)  & \cos (2 \pi \lambda R) 
\end{array}
\right)
\left(
\begin{array}{c}
Z \\
Y
\end{array}
\right) \ , \label{t027}
\ea
which is nothing but a two-dimensional rotation ${\cal R}_{\theta}$  in the
$(Z,Y)$ space, with an angle $\theta = 2 \pi \lambda R$. 
The orientifold  identification
$y\rightarrow -y$ is mapped in terms of the coordinates $(Z,Y)$
to a parity $\Pi_Y$ times a rotation ${\cal R}_{2 \omega}$ with an angle 
$2 \omega$ 
\be
\Omega' \ = \ \Omega \ \Pi_Y \ {\cal R}_{2 \omega} \ . \label{t028}
\ee

The metric (\ref{t24}) and the orbifold and orientifold operations
(\ref{t027}), (\ref{t028}) allow a physical 
interpretation of our configuration in the $(Z,Y)$ coordinates. 
Indeed, the fixed points of the two orientifold operations are
\ba
\Omega'~: \  Y &=& \tan \omega \ Z \ , \nonumber \\
\Omega'~ {\cal R}_{2 \pi \lambda R} : \  Y &=& \tan (\pi 
\lambda R + \omega ) \ Z \ . \label{t029}
\ea
Consequently, the orientifolds and
branes located at the origin have an angle $\theta_0 = \ \omega $
with respect to the $Z$ axis,
whereas the  branes and orientifolds at $y=\pi R$ are at an angle 
$ \theta_1 = \ \pi\lambda R + \omega $. In the new coordinate system,
the interpretation of (\ref{t25}) is quite simple. 
It corresponds to the dilaton NS-NS tadpole condition obtained from the
action (\ref{t1b}), when we correctly take into account in the
Born-Infeld action the different rotation of the two sets of D-branes
and O-planes.   
Notice that the orbifold identification (\ref{t027}) imply that the
two-dimensional $(Z,Y)$ plane has singularities. For discrete values of
the rotation angle $\lambda = 1/(NR)$ or $M/(NR)$ with $M$ and $N$ coprime,
these singularities are of a conical
type and the resulting model before the orientifold operation (\ref{t029}) is 
the noncompact $C/Z_N$ orbifold model. 

\subsection{Freely-acting Lorentzian orbifold models}

It was shown in \cite{hp} that Lorentzian orbifolds are unstable,
since adding sources changes completely the spacetime which collapses
into a large black hole. It was also pointed out in \cite{hp,lms} that
making the orbifold boost to be freely-acting by combining it with a
shift in an additional coordinate can cure this problem in some cases.
We would like here to point out that such freely-acting boost operations 
can also be obtained from some specific nonsupersymmetric orientifold models.
For definiteness we restrict ourselves again to the case of the two-dimensional
boost. Let us start with Type IIB, orbifolded by $(-1)^F \delta$,
where $(-1)^F$ is the spacetime fermion number and $\delta$ a shift
acting simultaneously in two compact coordinates $y_1,y_2$
\be
\delta \ y_1 = y_1 + \pi R_1 \quad , \quad \delta \ y_2 = y_2 + \pi R_2
\ . \label{t040}
\ee
This orbifold breaks completely supersymmetry and generates a negative one-loop
bulk cosmological constant $\Lambda_1$, but has no fixed
points. We now orientifold by the operation $\Omega'= \Omega
\Pi_{y_1}$, where $\Pi_{y_1}$ is a parity in $y_1$. The fixed points under 
$\Omega'$ will generate O8 planes. By consistency, we must introduce
D8 branes in the theory\footnote{Perturbative orientifold models of
this type can be explicitly constructed \cite{gianfranco}.}. The
low-energy effective action is of the form (\ref{t1b}). By the
chain of arguments discussed in the previous paragraphs, the classical  
background of this model is equivalent to a two-dimensional orbifold
operation, supplemented by a shift 
\ba
T+X &\rightarrow & e^{2 \pi \lambda R} (T+X) \quad , \quad
T-X \rightarrow e^{-2 \pi \lambda R} (T-X) \ , \nonumber \\
y_2 &\rightarrow & y_2 + \pi R_2 \ , \label{t30} 
\ea
where $(T,X)$ are defined in terms of $(t,y_1)$ as in (\ref{t8}). 
In addition, the orientifold operation acting in the $(T,X,y_2)$
coordinates includes a boost, $\Omega' = \Omega \Pi_X {\cal K}_{2 \omega}$,
analogously to (\ref{t09}).

This type of smoothing of the singularity for the usual Milne space
turns out to be not sufficient to cure the instability problem
discussed in \cite{hp} and presumably also in the context discussed here.
However, in other examples, like the ``null-brane'' orbifold \cite{lms},
it does help. It would therefore be interesting to find
nonsupersymmetric vacua related in the sense described in the previous
section to the null brane orbifold \cite{lms}.

A similar construction can be applied to models with a background
depending on two space coordinates (\ref{t24}). This can help in order
to give a simple meaning to the case where the parameter $\lambda$ is
generic, by combining an irrational angle in the $(Z,Y)$ plane with a
translation along an additional circle. This actually defines a
Melvin-type model \cite{melvin}. There seems therefore to be a
surprising connection between nonsupersymmetric orientifolds 
with NS-NS tadpoles and Melvin type string models.
 
\section{Cosmological applications}
\begin{figure}
\vspace{6 cm}
\special{hscale=60 vscale=60 voffset=-15 hoffset=120
psfile=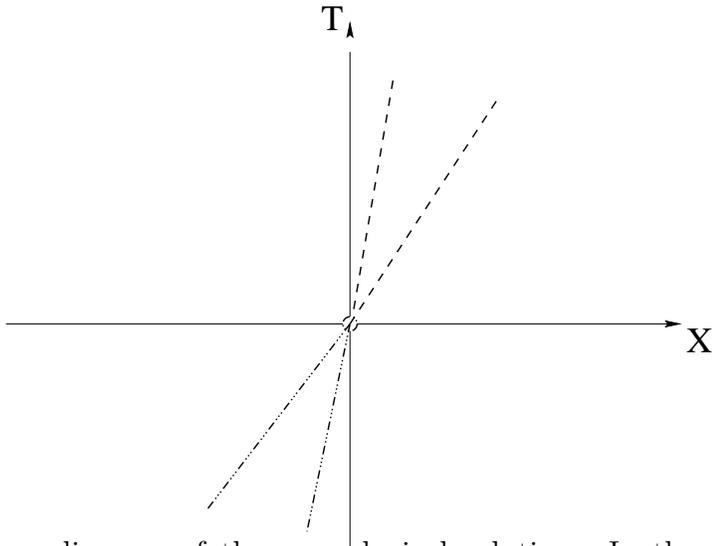}
\caption{The spacetime diagram of the cosmological solution. In
the region $T<0$ the time reversed evolution is drawn. The spacetime is
contained in the cone defined by the two orientifold planes, denoted by
dash-dotted lines.}
\end{figure} 
The spacetime diagram representing the solution (\ref{t9})
is depicted in figure 1. The bulk is the region between the two
dashed lines which represents the two moving boundaries.
At $T=0$ there is a big bang type singularity and the two boundaries
coincide. Near this region of spacetime the effective field
theory approach we are pursuing is not valid due to the higher
order $\alpha'$ corrections. Note however that the string coupling
is small in this region so that higher genus corrections are
expected to be negligible.
With the help of (\ref{t9}), the proper 
distance separating the
two fixed points under the orientifold involution
is easily calculated and is given in the Einstein frame by
\be
{8 \over 25{\kappa} \sqrt{\Lambda_e}}
\biggl[ [G_0 + 3{\kappa} \sqrt{\Lambda_e}  v_1 T]^{25 \over 24} -
 [G_0 + 3{\kappa} \sqrt{\Lambda_e} v_0 T]^{25 \over 24}  \biggr] \ , \label{t15}
\ee
where we recall that $v_0,v_1$ are the velocities of the two boundaries
(\ref{t12}), (\ref{t13}).

The distance becomes arbitrarily small for small $T$ signalling a
breakdown of the classical effective description.
 In the lower part of the diagram in
figure 1 we draw the time reversed solution which represent two
collapsing boundaries. Assuming, as in \cite{veneziano} and
\cite{epkyrotic} that string corrections allow a smooth
transition between the two branches, one suppresses the big bang
type singularity and gets an eternal spacetime.
This could offer a new perspective on the proposals
of \cite{veneziano}, \cite{epkyrotic}, since our string model
realizes a scenario similar to the one proposed there 
and hopefully the string corrections are easier to handle.

Since matter and Yang-Mills interactions are confined to the
boundaries, the relevant metric to consider, from the point of
view of an observer on the branes, is the induced one on the boundaries. 
Let us for simplicity suppose
that $\omega=0$. The metric induced at the origin $y=0=X$ is just a flat
metric, and the one induced at $y=\pi R$ is given by
\be
ds^2 =  
\biggl[ G_0+ 3{\kappa} sh (\pi \lambda R) \sqrt{\Lambda_e} \tilde T   
\biggr]^{1 \over 12} \biggl[  
\delta_{\mu \nu} dx^{\mu} dx^{\nu} -d\tilde T^2+ dX^2 \biggr]
\ , \label{t779}   
\ee
where we performed the change of coordinates 
$\tilde T= T / ch (\pi \lambda R)$. 
The geometry of spacetime  experienced by
an observer on the boundaries  does not 
necessarily reveal the spacetime singularities 
of the full ten dimensional geometry. This is particularly
obvious for the induced metric on $y=0$ which is completely flat
and does not ``see'' the  singularity at $X=T=0$.
On the other boundary, where the metric is (\ref{t779})
the observer experiences an expanding universe with
a singularity in the past at 
$\tilde T = - G_0/ (3{\kappa} sh (\pi \lambda R) \sqrt{\Lambda_e})$. 
However, this singularity
is just an illusion since the metric looses its validity before
that time at $\tilde T=0$, where the true singularity takes place.

We now turn to examine the backgrounds of different 
nonsupersymmetric models.
\section{Explicit string examples}
\subsection{The $SO(N)\times USp(32-N)$ model}

The model we consider here is an orientifold of the nine dimensional
Scherk-Schwarz deformation of Type IIA strings \cite{rohm}. There are
several consistent orientifold operations already considered in the
literature \cite{ads1}.  The example we are studying here is based on
the involution \cite{dm3} $\Omega' = \Omega (-1)^{F_L} \Pi_y$, where  $(-1)^{F_L}$
is the left world-sheet fermion number and $\Pi_y$ is the parity
in a compact coordinate $y$. The virtue of this projection, similar in
spirit with the one used in Type O orientifolds \cite{augusto2}, is
that it eliminates the closed string tachyon present in the Scherk-Schwarz
compactification. The fixed points of the orientifold operation are
$y=0$, containing an $O8_{+}$ plane with $(-16,-16)$ units of RR and
NS-NS charges and $y=\pi R$ containing an $\overline{O8}_{-}$
plane of charges $(-16,+16)$.  

As shown in Appendix B, RR tadpole cancellation in this model asks for a
net number of $N=32$
D8 branes placed in fixed positions on the compact coordinate $y$. A
nice feature of this model, which was one original motivation for
studying it, is that if all D8 branes are at the origin, the open massless spectrum 
is supersymmetric with gauge group $SO(32)$, while if all D8 branes
are placed in the other fixed point $y=\pi R$, the massless open
spectrum precisely coincides with the one of the 10d model
\cite{sugimoto}, with a gauge group $USp(32)$. Moving continously the
branes from one fixed point to the other interpolates therefore
continously between the supersymmetric $SO(32)$ and the
nonsupersymmetric\footnote{In this case, supersymmetry is however
nonlinearly realized in the open sector \cite{dm2}.}
$USp(32)$ Type I strings. Of course, this statement is strictly
speaking incorrect, since the closed (bulk) sector has softly
broken supersymmetry governed by the radius $1/R$, but for large
enough radius the main supersymmetry breaking appears due to the
simultaneous presence of $D8$ branes and $\overline{O8}_{-}$ planes,
which breaks supersymmetry at string scale, if they are in top of each
other, or if not at a scale proportional to the distance between them. 
 
One of the puzzles of this model which motivated our investigation is
the interaction pattern between D branes and O planes of this
model. In flat space, the $O8_{+}$ plane at $y=0$ and the $\overline{O8}_{-}$ 
at $y=\pi R$ repel each other, a rather surprising feature, whereas
O-planes and  anti O-planes of the same type
attract each other. Moreover, the D-branes do not interact with the
$O8_{+}$ planes by supersymmetry, whereas they are attracted by the $\overline{O8}_{-}$
planes. Consequently, if we start with an initial configuration of D8
branes on top of $O8_{+}$ planes, with massless supersymmetric spectrum
and gauge group $SO(32)$, the dynamics of the system seems to push the D8
branes to move towards $y=\pi R$ and the final state of the system is
the $USp(32)$ nonsupersymmetric string ! 
 
In analogy with Section 2, we place N D8 branes at $y=0$ and the rest 32-N
at $y= \pi R$. The effective action of the system is (\ref{t1b}), 
with $\beta_0=(16-N) {\kappa}^2 T_8$, $\beta_1=(48-N) {\kappa}^2 T_8$ and
$q_0 = T_0$. There is a bulk one-loop cosmological constant,
which is small in the large-radius limit $R >>
\sqrt{\alpha'}$, such that the effective cosmological constant is
positive $\Lambda_e >0$ and we are in the case {\bf i)} of section 2.
To start with, we neglect the one loop cosmological constant.
The equations are readily solved and the solution is (\ref{t3}) with $\omega=0$.
In this case, according to (\ref{t12}), the bounadry at the
origin has zero velocity $v_0=0$, whereas (\ref{t14}) fixes the boost parameter
$\lambda$ by the equation $\beta_0 \ ch (\lambda \pi R) \ = \ \beta_1$.
This relation shows that solutions
with the symmetry displayed in (\ref{t2}) exist
provided that the number of branes at the origin is less than $16$
($\beta_0>0$) and therefore
the tension due to the orientifolds and the branes at the origin must
be negative.  

It turns out that
the sign of the one-loop cosmological constant is crucial.
Indeed, since $T_0=q_0$, eq. (\ref{eq}) implies that there are no solutions
for a positive one loop cosmological constant, no matter how small
it is. Since supersymmetry is broken in the closed sector by a Scherk-Schwarz mechanism, the
resulting one-loop cosmological constant is negative
and (\ref{eq}) does have solutions. 

 We leave for future work a
complete analysis of the dynamics of a probe brane in this
background. Here we just note that, since the bulk is supersymmetric, a
particular solution for the position of the test brane is $X = {\rm
cst.}$. In this case, since the one-loop bulk cosmological constant is
small, the velocity of the supersymmetric boundary at $y=0$ is much smaller than the positive
velocity of the nonsupersymmetric boundary at $y=\pi R$. 
Therefore, for this particular solution the test brane will stay closer
to the supersymmetric boundary, despite the naive flat space arguments presented above. 
 
\subsection{Other models}

The solutions we found in Section 2 can be
used for other string models with broken supersymmetry.

An interesting model to analyse is 
the nine dimensional model with 16 $O8_{+}$ planes at the
origin and 16 anti-orientifold ${\overline O8}_{+}$ planes at the other fixed point
$y=\pi R$. In fact, this model is a different orientifold \cite{ads1}
of the same Scherk-Schwarz deformation of the type II string.
The model do not need branes for consistency, since the RR charges
add up to zero. In addition, the tachyon is not removed by the orientifold
projection for any values of the radius. Let us however add an equal number $N$
of brane-antibrane pairs, the branes being placed at the origin and the
antibranes at $y=\pi R$, in order to avoid the occurence of open-string tachyons.
For this configuration, we obtain
\ba
T_0 &=& (N-16) T_8 \quad , \quad T_1 = (N-16) T_8 \ , \nonumber \\
q_0 &=& -(N+16) T_8 \quad , \quad q_1 = (N+16) T_8 \ . \label{o1} 
\ea  
The bulk one-loop cosmological constant in this model is small in the
large radius limit and therefore the conditions (\ref{condi2}) are satisfied. 
We obtain therefore solutions of the type (\ref{t24}), with the
parameter $\lambda = 1/R$. In the $(Z,Y)$ coordinate system, the
orbifold identification (\ref{t027}) becomes trivial and therefore we
get a regular two-dimensional noncompact plane.  

Since the brane-antibrane pairs are not needed by the consistency of the
theory, we can also analyse the particular case $N=0$.
The classical background can be easily deduced from 
section 2. In fact this case corresponds to $\beta_0 = - \beta_1$
and $\beta_0 > 0$. If we take into account the negative one loop
cosmological constant we can get solutions of the form (\ref{t18}), but
only provided that the effective cosmological constant $\Lambda_e$ is
negative. This correponds however to a radius $R$ of the order
the string scale, where our effective theory analysis is not reliable.

Another model which can be similarly analysed is the 
T-dual of the $USp(32)$ nonsupersymmetric string compactified on a circle.
The model has 16 $O8_{-}$ planes at each fixed point and 
$32$  ${\overline D8}$ branes which for simplicity can be distributed
among the fixed points, $N$ at $y=0$ and $32-N$ at $y=\pi R$.
Here, the one loop bulk cosmological constant is
zero. A quick inspection of (\ref{eq}) supplemented with the boundary
conditions (\ref{bor}) shows that there are no solutions for any values
of N. 

We can also analyse the case of the supersymmetric Type I' string
supplemented by a certain number $N$ of brane-antibrane pairs. In order
to avoid open-string tachyons, we place all
$32+N$ D8 branes at the origin and all $N$ ${\overline D8}$ antibranes at $y
= \pi R$. In this case,  we have
\ba
T_0 &=& (N+16) T_8 \quad , \quad T_1 = (N-16) T_8 \nonumber \\
q_0 &=& (N+16) T_8 \quad , \quad q_1 = -(N+16) T_8 \label{o3}
\ea  
and the bulk one-loop cosmological constant is zero.
The conditions (\ref{condi}) in this case are violated, while
(\ref{condi2}) are satisfied. Therefore, a solution exists depending
on two space coordinates $(y,z)$ given explicitly by (\ref{t24}), with
$\omega=0$ and $\lambda$ determined by $\sin (\pi \lambda R) = (16-N)/(N+16)$. 


\vskip 24pt

\noindent
{\bf Acknowledgements.} 
We are grateful to C. Charmousis, G. D'Appollonio and G. Pradisi for
useful discussions. E.D. was
supported in part by the RTN European Program HPRN-CT-2000-00148. 
\appendix
\section{Appendix: Equations of motion and more solutions}

In this appendix we look for solutions to the 
equations of motion (\ref{e1}-\ref{e4}) supplemented with the
boundary conditions (\ref{bor}). We will be able to find all the
solutions with $k=0$ and special solutions with $k=-1$.

The combination (\ref{e3})-(\ref{e4})+2/5(\ref{e1})
using the relations (\ref{an}) gives
\be
\ddot {\phi} + 8{\dot A} {\dot \phi}
 +140 k e^{2b}=0.
\ee
>From (\ref{fac}) we get $\dot A$ in terms of $F$ and $G$ and their
derivatives. The previous equation becomes
\be
\ddot {\phi} + {1 \over 3}\left(
{{\dot G} \over {G+F}}+{\dot b}+{5 \over 4} {\dot\phi}\right) 
\dot{\phi} +140 k e^{2b}=0 \ . \label{combi}
\ee
In this equation only $F$ depends on $y$. If we assume that
at least one among $\beta_0$ and $\beta_1$ is not zero than 
due to (\ref{bor}) $F$ cannot be constant. This implies that
$\dot{\phi} \dot{G}=0$. We have to consider two cases: \
a) $\dot{\phi}=0$ and \ b) $\dot{G}=0$.

\begin{itemize}

\item{\bf Case a}: Equation (\ref{combi})  with $\dot{\phi}=0$ 
gives $k=0$, so only flat eight dimensional spaces are possible in
this case. The combination (\ref{e2})+(\ref{e4}) gives
\be
4\ddot{b}+(\dot{b} \dot{G}+\ddot{G})(F+G)^{-1}=0 \ .
\ee
This equation gives $\ddot{b}=0$ and $\dot{b} \dot{G}+\ddot{G}=0$,
so we have $\dot{b}=\lambda$, $\lambda$ being a constant and
\be
\lambda \dot{G}+\ddot{G}=0 \ . \label{inte1}  
\ee
Next, consider the equation (\ref{e3}). After using (\ref{inte1}),
it gives
\be
(\dot{G} +\lambda G)^2- 
(F^{\prime 2}-\lambda^2 F^2-9\kappa^2 \Lambda_e)+
2\lambda F(\dot{G}+\lambda G) = 0 \ . \label{inte2}
\ee 
This can hold provided that
\be
 \dot{G}+\lambda G=0 \quad , \quad 
F^{\prime 2}-\lambda^2 F^2 = 9 \kappa^2 \Lambda_e \ .
\ee
We have obtained all the equations (\ref{eq}) and (\ref{eq1}).
It is now possible to verify that the other equations are
identically satisfied. This case leads to solutions where the dilaton
never diverges and they are smoothly connected to
supersymmetric solutions in the limit of zero boost (rotation) and was
discussed in great detail in the text.
 
\item{\bf Case b}: Since $\dot{G}=0$, we can assume without loss
of generality that $G=0$ because an eventual constant $G$ can be
absorbed into $F$. Equation (\ref{combi}) now becomes
 \be
 \ddot {\phi} + {1 \over 3}\left(
 \dot{b}+{5 \over 4} \dot{\phi}\right) 
\dot{\phi} +140 k e^{2b}=0 \ . \label{combi2}
\ee
In this case (\ref{fac} implies 
\be
\dot{A}={1 \over 24}\left(
\dot{b}+{5 \over 4} \dot{\phi}\right) \quad , \quad
A^{\prime}={F^{\prime}\over 24F} \ ,
\ee
so that ${\dot A}$ ($A^{\prime}$) depends only on $t$ $(y)$.
The four equations (\ref{e1}-\ref{e4})
have thus the form
\be
{\cal F}_i(y) \ + \ {\cal G}_i(t)=0 \quad , \quad i=1,\dots 4
\ee
for some functions ${\cal F}_i$ and ${\cal G}_i$.
This implies that both must be constants, that is
\be
{\cal F}_i=a_i \quad , \quad {\cal G}_i=-a_i \ ,
\ee
where $a_i$ are four constants. The $y$ dependent part of the
equations is quite simple. For instance equation (\ref{e3})
gives
\be
F^{\prime 2}-9a_3F^2 = 9 \kappa^2 \Lambda_e \ .
\ee
Inserting this equation in the other ones ${\cal F}_i$
determines the constants
$a_1$, $a_2$ and $a_4$ in terms of $a_3$. The remaining
equations read
\ba
&& (\ddot{A}+\ddot{b})+8\dot{A}(\dot{A}+\dot{b})={25 a_3 \over
8} \ , \label{ee1} \\
&& \ddot{A}+8\dot{A}^2+\dot{A}\dot{b}+{1 \over 8}
\ddot{b}+{49 \over 8} k e^{2b}={a_3 \over 2} \ , \label{ee2}\\
&& \ddot{A}+8\dot{A}^2+7k e^{2b}={a_3 \over 8} \ , \label{ee3} \\
&& 32\times 7 \dot{A}^2-\dot{b}^2+48\dot{A}\dot{b}
+7\times 25 k e^{2b}={25 a_3 \over 2} \ . \label{ee4}
\ea
Notice that all the equations are not independent since for
instance  (\ref{ee1})-8(\ref{ee2})+7(\ref{ee3})
is identically zero.

Let us first consider the $k=0$ case.
If we define $h=\exp{((b+5\phi/4)/3)}$, then the
equation (\ref{ee3}) reads
\be
\ddot{h}-a_3h=0 \ .
\ee
A direct consequence of this equation is
\be 
\dot{h}^2-a_3h^2=E \ ,
\ee
where $E$ is a constant. Equation (\ref{combi2}) now can be solved 
as
\be
\dot{\phi}={c \over h} \ ,
\ee
where $c$ is a constant. Equation (\ref{ee1}) and (\ref{ee2})
are identically satisfied and equation (\ref{ee4}) gives $E=c^2/8$.
If $a_3= \lambda^2/9 > 0$, then the solution reads
\be
h(t)= {3 c \over {2\sqrt{2}\lambda}} \ sh({\lambda (t-t_0) \over 3}) \ .
\ee
The spacetime metric and the string coupling are given by
\ba
ds^2 &=&  \biggl[ [sh {\lambda
(t-t_0) \over 3}]^3 F(y) \biggr]^{1 \over 12} \times
\nonumber \\ 
&& \biggl[ \delta_{\mu \nu} dx^{\mu} dx^{\nu} + [sh {\lambda
(t-t_0) \over 3}]^6  [th {\lambda (t-t_0) \over 6}]^{+5 \sqrt{2}} (-dt^2+dy^2) \biggr]
\ , \nonumber \\
e^{\Phi} &=& e^{-5 b_0/6 - \phi_0/24} [sh {\lambda (t-t_0) \over 3}]^{-5/2}
[th {\lambda (t-t_0) \over 6}]^{\sqrt{2}} F(y)^{-5/6}  
 \ , \label{a1}   
\ea
where $t_0$ is an integration constant and the parameters $\lambda$ and
$\omega$ are determined by (\ref{t7}). The physical time-region can be
taken for example between $t_0$ and $\infty$. 
The final solution depends, as in the solutions discussed in the text,
on the sign of the effective cosmological constant.
For the three different values of $\Lambda_e$, the $y$-dependent
function $F(y)$ is
\ba
&& \Lambda_e > 0 \quad , \quad F(y) = {3 \kappa \sqrt{\Lambda_e} \over
\lambda} sh (\lambda |y| + \omega) \ , \nonumber \\
&& \Lambda_e < 0 \quad , \quad F(y) = {3 \kappa \sqrt{\Lambda_e} \over
\lambda} ch (\lambda |y| + \omega) \ , \nonumber \\
&& \Lambda_e =0 \quad , \quad F(y) = F_0 e^{\lambda |y|} \ , \label{a01}
\ea
where the parameters $\lambda,\omega$ are determined by (\ref{t7}) if 
$\Lambda_e > 0$, by (\ref{t17}) if $\Lambda_e < 0$ and by (\ref{t22}) if
$\Lambda_e =0$.

These solutions have
large string coupling close to the big bang singularity. Moreover, they
are not smoothly connected in the $\lambda=\omega=0$ limit to the
supersymmetric solution \cite{pw}.
 
In the case $a_3 = - \lambda^2/9 < 0$, the solution is
\be
h(t)={3 c \over {2\sqrt{2}\lambda}} \ \sin({\lambda (t-t_0) \over 3}) \ .
\ee
The solution exists only when the effective cosmological constant
is positive. The final solution in this case is
\ba
e^{\Phi} &=& e^{-5 b_0/6 - \phi_0/24} 
[\cos {\lambda (t-t_0) \over 3}]^{-5/2}
[ctg ({\lambda (t-t_0) \over 6}+{\pi \over 4})]^{2\sqrt{2}} 
[{3 {\kappa} \sqrt{\Lambda_e}
\over \lambda} \sin (\lambda |y|+ \omega)]^{-5/6}  
 \ ,\nonumber \\
ds^2 &=&  \biggl[ {3 {\kappa} \sqrt{\Lambda_e} \over \lambda} [\cos {\lambda
(t-t_0) \over 3}]^3 \sin (\lambda |y|+\omega) \biggr]^{1 \over 12} \times
\nonumber \\ 
&& \biggl[ \delta_{\mu \nu} dx^{\mu} dx^{\nu} +  |\cos {\lambda
(t-t_0) \over 3}|^{2/3}  |ctg ({\lambda (t-t_0) \over 6}+{\pi \over 4})|^{+5
\sqrt{2}} (-dt^2+dy^2) \biggr] \ , 
 \label{a2}   
\ea
where the parameters $\lambda$ and $\omega$ are determined here by (\ref{t23}).

\end{itemize}
 
We were able to find particular curved solutions, $k\neq 0$ of the system
(\ref{ee1}-\ref{ee4})for $k=-1$. The solution has a constant $b$ and $\dot{\phi}$:
\be
\dot{\phi}^2 = 9\times 16 a_3 \quad , \quad e^{2b} = {3 \over 7}a_3 \ .
\ee
The solution, which exists only for $a_3>0$, is explicily given by
\ba
ds^2 &=& e^{5 \phi_0 /4 } [ e^{5\lambda t} F(y)]^{1/12}  
\left[ \left(1-{ x^2 \over
4}\right)^{-2} \delta_{\mu \nu} dx^{\mu} dx^{\nu} + 
{\lambda^2 \over 21} (-dt^2 + dy^2)\right],\\
e^{\Phi}&=& e^{- \phi_0 /24} \ F(y)^{-5/6} e^{-\lambda t/6} \ , \label{a3}
\ea
where for the three different values of $\Lambda_e$, the $y$-dependent
function $F(y)$ is given by (\ref{a01}).

Contrary to the flat solutions discussed in the text, all the backgrounds
discussed in the Appendix cannot be transformed by a change of coordinates
into static ones. 

There are also static solutions, depending on coordinates $(z,y)$, with
$z$ being parallel to the branes. There are also two cases to consider,
and the analog of case {\bf a} was already discussed in the text. In
case {\bf b}, eqs. (\ref{ee1})-{\ref{ee4}) still hold provided we flip
the sign in the right hand side of the equations. Here $k=1$ corresponds to
the dS eight-dim. space, whereas $k=-1$ corresponds to
the AdS eight-dim. space. The solutions have a form similar to the one
appearing in (\ref{a1}) and (\ref{a2}) with the time-depence replaced
by a $z$-dependence such that hyperbolic functions are replaced by
trigonometric functions and vice-versa.
  
\section{Appendix: Nine dimensional nonsupersymmetric orientifold}

In this appendix we give the technical details
for the string model {\cite{dm3} considered in section 4.
It is simpler to construct its T-dual version. 
We start from a Scherk-Schwarz deformation of IIB
which is obtained by modding IIB on a circle of radius $r$
by $(-1)^{F}\times
\sigma$, where $F$ 
is the spacetime fermion number and 
$\sigma$ acts on a circle $S^1$ 
as an asymmetric shift $y_R\rightarrow y_R+\pi r$,
$y_L\rightarrow y_L-\pi r$.

The resulting torus partition function is:
\ba
 T&=&\frac{1}{2} \left\{
   |V_8-S_8|^2\Lambda_{m,n}+|V_8+S_8|^2(-1)^n
  \Lambda_{m,n}\right\} \nonumber \\
&+&\frac{1}{2} \left\{
|O_8-C_8|^2\Lambda_{n,m+\frac{1}{2}}+
|O_8+C_8|^2(-1)^n\Lambda_{m+\frac{1}{2},n}
\right\}, \nonumber \\ 
\ea
where the $SO(8)$ characters are defined as:
\be 
O_8=\frac{\theta_3^4+\theta_4^4}{2\eta^4}, \ \ \
V_8=\frac{\theta_3^4-\theta_4^4}{2\eta^4},\ \ \ S_8=\frac{\theta_2^4-\theta_1^4}{2\eta^4},\ \ \ C_8=\frac{\theta_2^4+\theta_1^4}{2\eta^4}\ee

with $\theta _i$ the Jacobi functions and $\eta$ the Dedekind 
function
and
\be
\Lambda_{m,n} = \sum_{m,n} q^{{\alpha' \over 4} ({m \over r}+{n r \over \alpha'})^2}
{\bar q}^{{\alpha' \over 4} ({m \over r}- {n r \over \alpha'})^2} \ . \label{n2}
\ee
The resulting model interpolates between the IIB theory in
the limit $r\rightarrow 0$ and the 
0B theory in the limit $r\rightarrow \infty$.

After performing a rescaling of the radius 
$r\rightarrow 2r$ the torus
 amplitude reads: 
\ba 
T &=& \left \{ (|V_8|^2+|S_8|^2)\Lambda_{2m,n}+
 (|O_8|^2+|C_8|^2)\Lambda_{2m+1,n}\right \} \nonumber \\
&-& \left \{(V_8\overline{S}_8+S_8\overline{V}_8)\Lambda_{2m,n+\frac{1}{2}}+ 
 (O_8\overline{C}_8+C_8\overline{O}_8)\Lambda_{2m+1,n+\frac{1}{2}}\right\}
\ . \nonumber 
\ea
Next, we consider the  orientifold obtained by
gauging the discrete symmetry 
$ \Omega ' = \ \Omega (-1)^{F_L} \ $, 
where $\Omega$ is the standard worldsheet parity operator
and $(-1)^{F_L}$ is the worldsheet fermion number.
The resulting Klein bottle is given by:
\ba 
K &=& \frac{1}{2}\left\{ (V_8-S_8)P_{2m}-
(O_8-C_8)P_{2m+1}\right \}, \nonumber
\ea
with 
$P_m= \sum_m q^{\alpha' m^2 \over 4 R^2}$.
The tachyon is antisymmetrised in the Klein bottle and removed
from the spectrum. Notice that the limit where the radius goes
to infinity reduces to the closed sector of the O'B model.

The tadpoles are obtained from the transverse channel amplitude
\ba 
\widetilde{K}
= \frac{2^5}{2}(V_8-S_8)W_n-(V_8+S_8)(-1)^n W_n
= \frac{2^5}{2}(V_8W_{2n+1}-S_8W_{2n}) \ ,  
\ea
where
$W_n = \sum_n q^{n^2 R^2 \over \alpha'}$.

We can see that there is no NS-NS tadpole, but there is a R-R tadpole.
The cancelation of the tadpoles will require the introduction of $D9$
branes.
In a T-dual spacetime interpretation, 
the model contains 16 $O8_{+}$ and
16 $O{\bar 8}_{-}$ planes.

The open sector amplitudes for the D9 branes are
\ba
A = \biggl\{ {N_1^2+N_2^2 \over 2}  \sum_m q^{\alpha' m^2 \over R^2}
+ N_1N_2  \sum_m q^{\alpha' (m+1/2)^2 \over R^2}  \biggr\} (V_8-S_8) \ ,
\nonumber \\
M = -{N_1-N_2 \over 2}  \sum_m (-1)^m q^{\alpha' m^2 \over R^2} V_8
+ {N_1+N_2 \over 2}  \sum_m q^{\alpha' m^2 \over R^2} S_8 \ , \label{n5}
\ea
The RR tadpole cancellation asks for $N_1+N_2=32$ and
the resulting gauge group is $SO(N_1) \times USp(N_2)$.
In the T-dual version we have $N_1$ D8 branes at the origin and
$N_2=32-N_1$ branes at the other fixed point.
When $N_2=0$ the open spectrum is that of the supersymmetric 
type I, whereas the $N_1=0$ open spectrum is
 the one of the  $USp(32)$ non-susy model.
 
Notice that in the limit $R\rightarrow \infty$
the RR tadpole cancellation condition is modified
and the resulting gauge group is $U(32)$.


\end{document}